\documentclass[aps,prl,floatfix,twocolumn,superscriptaddress,reprint]{revtex4-2}
\usepackage{amsmath}
\usepackage{amssymb}
\usepackage{eurosym}
\usepackage{amsbsy}
\usepackage{subfigure}
\usepackage{comment}
\usepackage{color}
\usepackage{mathrsfs}
\usepackage{amsfonts}
\usepackage{xspace}
\usepackage{latexsym,epsfig,graphicx}
\usepackage{epstopdf}
\usepackage{tabularx}
\usepackage{longtable}
\usepackage{bm, amsfonts, braket}
\usepackage{amsthm}
\usepackage{titlesec}

\usepackage[normalem]{ulem}
\usepackage[version=4]{mhchem}
\usepackage{multirow}
\usepackage[mathlines]{lineno}
\usepackage{dcolumn}
\usepackage[colorlinks=true, pdfstartview=FitV, linkcolor=blue, citecolor=blue, urlcolor=blue]{hyperref}
\usepackage{bm}

\newcommand\wrapped[1]%
  {\renewcommand\arraystretch{1}%
   \begin{array}{@{}l@{}}#1\end{array}%
  }
\usepackage{xcolor}

\begin{document}
\preprint{APS/123-QED}

\title{A geometry-originated universal relation for arbitrary convex hard particles}
\author{Yuheng Yang}
\affiliation{Key Laboratory of Artificial Micro- and Nano-structures of Ministry of Education and School of Physics and Technology, Wuhan University, Wuhan 430072, China}
\author{Duanduan Wan}
\email[E-mail: ]{ddwan@whu.edu.cn}
\affiliation{Key Laboratory of Artificial Micro- and Nano-structures of Ministry of Education and School of Physics and Technology, Wuhan University, Wuhan 430072, China}
\date{\today}

\begin{abstract}

We have discovered that two significant quantities within hard particle systems: the probability of successfully inserting an additional particle at random and the scale distribution function, can be connected by a concise relation. We anticipate that this relation holds universal applicability for convex hard particles. Our investigations encompassed a range of particle shapes, including one-dimensional line segments, two-dimensional disks, equilateral and non-equilateral triangles, squares, rectangles, and three-dimensional spheres. Remarkably, we have observed a close alignment between the two sides of the relation in all cases we examined. 
Furthermore, we show that this relation can be derived from the fundamental thermodynamic relation that connects entropy, pressure, and chemical potential.
Our study unveils a geometrically rooted relation that underpins essential thermodynamic relations, shedding light on the intricate interplay of geometry and thermodynamics in hard particle systems.

\end{abstract}


\maketitle


Hard particle systems, where particles interact through a hard-core potential, are ``pure" systems in the sense that the potential energy is zero for all permissible configurations. In equilibrium, at a constant temperature $T$, a system with a fixed volume minimizes its Helmholtz free energy $F$, defined as $F = U - TS$, where $U$ is the internal energy and $S$ is the entropy. In the case of hard particle systems, since the kinetic energy is a constant and can be ignored at a fixed temperature, the equilibrium phase is determined solely by entropy \cite{Manoharan2015}. Despite the apparent simplicity, rich complexity can arise in hard particle systems, depending only on particle shape and packing fraction (e.g., Refs.~\cite{Torquato2009,  Haji-Akbari2009, Marechal2010, Gang2011, Agarwal2011,  Damasceno2012, Ni2012, Smallenburg2012, Avendano2012, Gantapara2013,  Bernard2011_melting, Anderson2017, Lei2018_helix, Klotsa2018, Wan2019, Wan2021}). For example, Haji-Akbari \textit{et al.} revealed that hard tetrahedra can form a dodecagonal quasicrystal within a specific packing fraction range \cite{Haji-Akbari2009}. Additionally, Damasceno and collaborators investigated the self-assembly of 145 convex polyhedra and observed a wide range of structures, including $\gamma$-brass, which possesses 52 atoms per unit cell and emerges from truncated dodecahedra \cite{Damasceno2012}. Furthermore, when a hard particle system is subjected to geometrical confinement, additional intriguing structures and phases can emerge, adding further depth to the exploration of these systems (e.g., Refs.~\cite{Carlsson2012_topology, Blair2012_squares, Teich2016, Sitta2018,  Wan2018,  Wan2022, Mughal2012_spheres, Jin2020_spheroids, Wan2023_chiral, Mbah2023}).

Two significant quantities are associated with the statistical properties of hard particle systems. One is the probability of successfully inserting an additional particle at random. Suppose there are already $(i-1)$ particles present with random positions and orientations, the probability of successfully inserting the $i$th particle, denoted as $P_{i}$, depends on the configuraiton of the previous $(i-1)$ particles. Interestingly, except for a constant coefficient, the natural logarithm of $P_{N+1}$ provides the chemical potential of a $N$-particle system \cite{Widom1963, Frenkel1996}. The second essential quantity is the scale distribution function $\tilde{s}(x)$, which characterizes the probability distribution of particle overlap with their nearest neighbors. For any given particle $i$, there exists a minimal value of $x$, leading to a maximal factor $(1-x)$. This maximal factor will result in an overlap between particle $i$ and its nearest neighbor if their separation becomes $(1-x)$ times their original value, while particle orientations remain unchanged. The scale distribution function is intimately related to the pressure of a hard particle system \cite{Eppenga1984,Brumby2011,Anderson2016}. 
Despite their distinct origins, both of these quantities provide insights into the degree of packing density within a system, and their values are influenced by the shape of the particles.

In this study, We have uncovered a profound connection between these two quantities in a monodisperse hard particle system, which we express through a concise equation:
\begin{equation}
\lim_{\substack{N \to \infty \\ N/V =\rho}} \, \frac{1}{N} \ln (\prod_{i=1}^{N} \frac{P_{N+1}}{P_{i}} ) = -\frac{\tilde{s}(0+) }{2d}.
\label{equ_relation} 
\end{equation}
Here $N$ is the total number of particles, $\rho$ is the density of particle number, $V$ is the system volume, determined once $N$ and $\rho$ are established. Both $P_i$ and $P_{N+1}$ are calculated while keeping the system volume $V$ constant. $\tilde{s}(0+)$ is the one-sided limit of the scale distribution function $\tilde{s}(x)$, as $x$ approaches zero, for this $N$-particle system. $d$ is the dimensionality of the system. 
In the thermodynamic limit, where the number of particles tends to infinity while maintaining a constant density ($N \to \infty$ with $N/V = \rho$), this relationship is rigorously valid. We anticipate that this relationship holds universally for convex hard particles of arbitrary dimensions and shapes.

\begin{figure}
\includegraphics[width=\columnwidth]{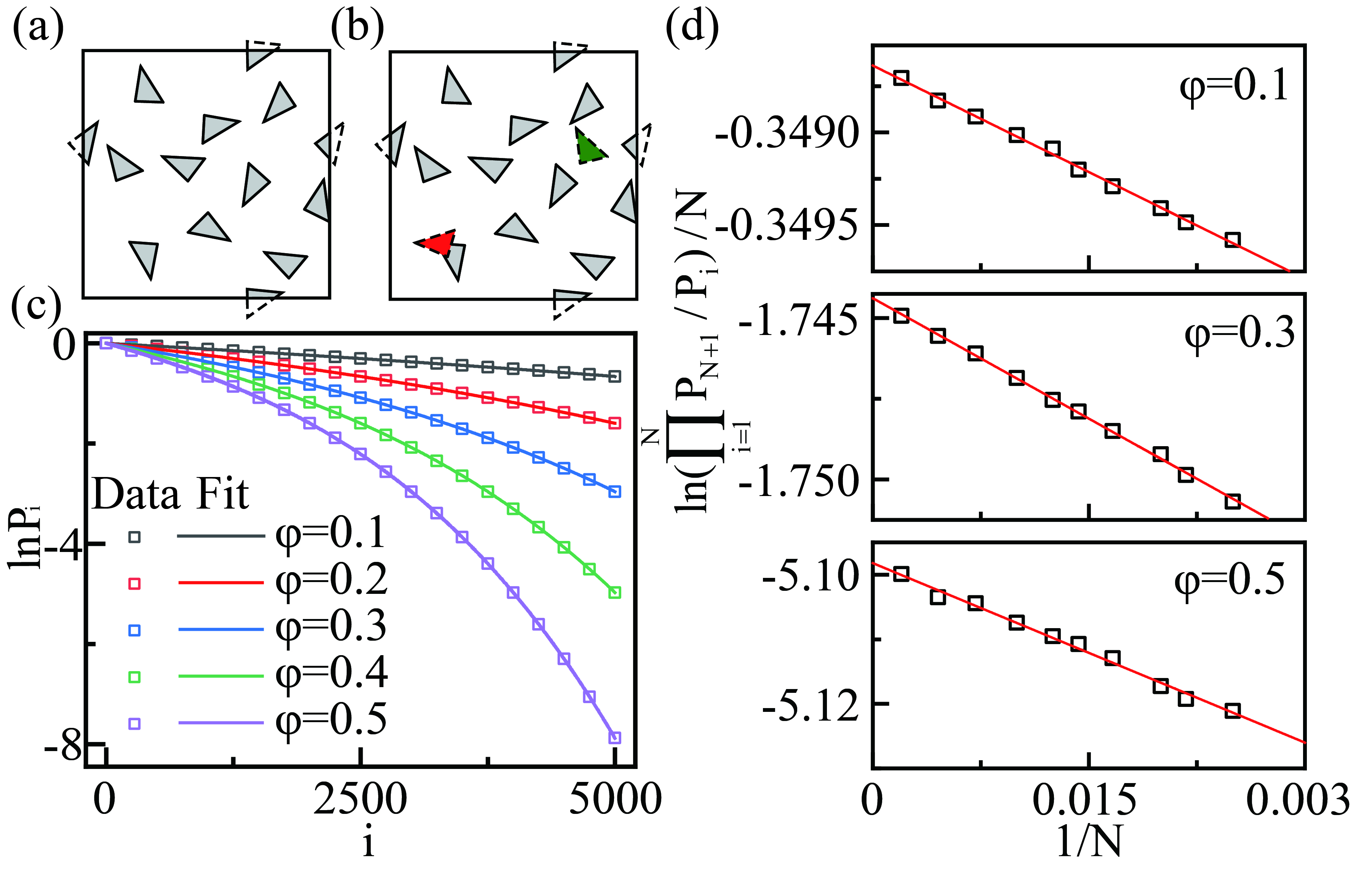} 
\caption{(Color online) (a) A two-dimensional non-equilateral triangle system with side lengths in a 4:5:6 ratio (depicted in light grey) in a periodic box. (b) The insertion of the next triangle can either be considered a success (shown in green) or a failure (indicated in red) depending on whether it overlaps with a previously placed triangle. (c) Plot of $P_{i}$ as a function of $i$ at selected values of $i$ for a system with $N=5000$. Curves are fourth-order polynomial fits to the data points. (d) Plot of $\ln (\prod_{i=1}^{N} P_{N+1}/P_{i} )/N$ as a function of $1/N$. Lines are linear fits to the data points. } 
\label{fig_triangle}
\end{figure}

\begin{figure}
\includegraphics[width=\columnwidth]{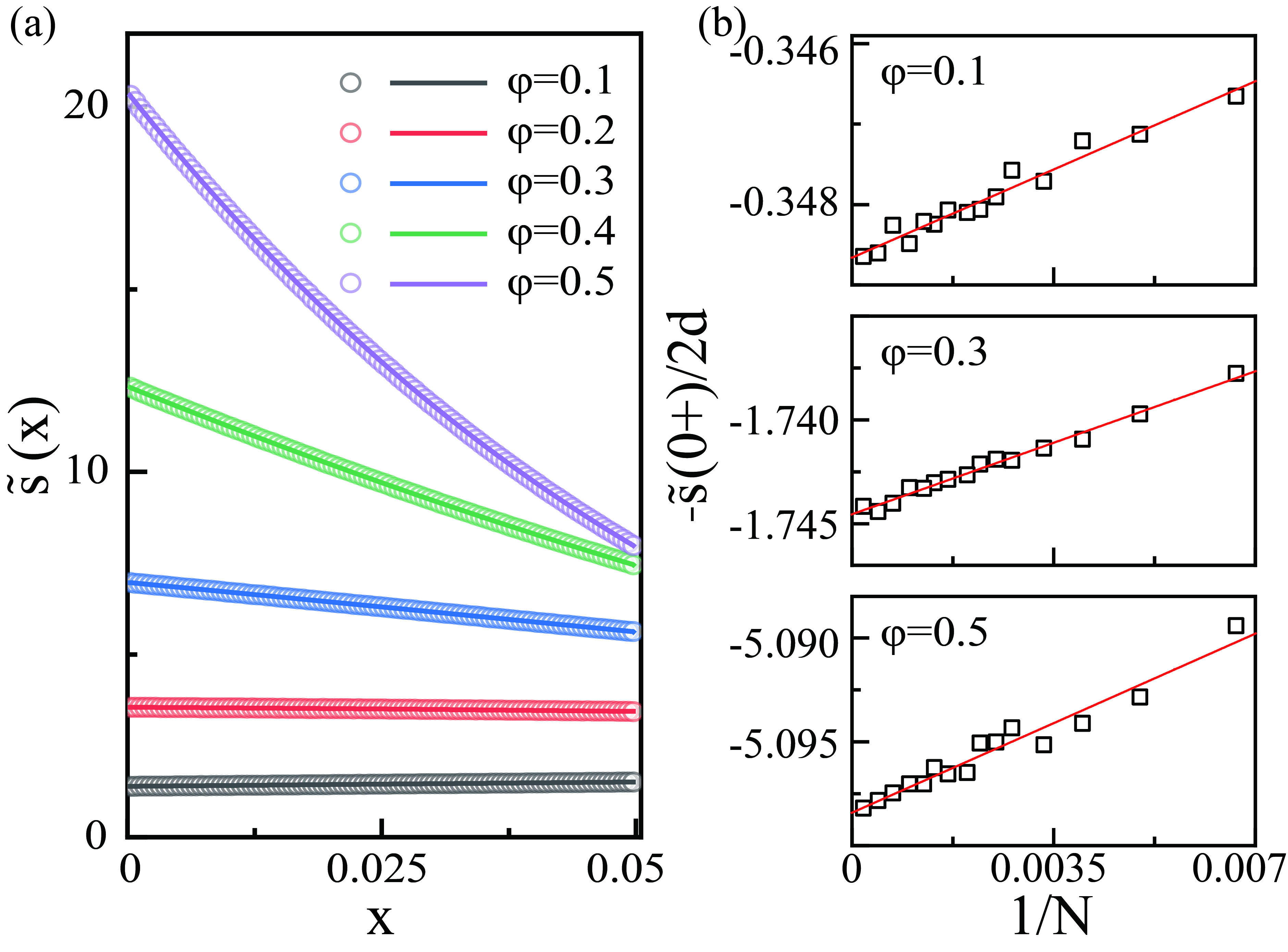} 
\caption{(Color online) (a) Plot of $\tilde{s}(x)$ as a function of the scale factor $x$ for the two-dimensional non-equilateral triangle system with $N = 5000$. Curves are
fourth-order polynomial fits to the data points. (b) Plot of $\tilde{s}(0+)$ as a function $1/N$. Lines are linear
fits to the data points. } 
\label{fig_sdf}
\end{figure}

To assess the validity of Eq.~(\ref{equ_relation}), we initiated our investigation with a system comprising $N$ identical hard particles confined within a cubic box of length $L$ and volume $V=L^{d}$, where $d$ denotes the dimensionality of the system. We conducted our simulations utilizing the hard particle Monte Carlo module implemented in HOOMD-blue \cite{Glaser2015, Anderson2020}, with the application of periodic boundary conditions.
For the sake of clarity, we focus on a two-dimensional system in this description, although analogous procedures were applied to systems of different dimensions. In the two-dimensional context, with a fixed number of particles $N$ and a packing fraction $\varphi$, we maintained the box area at $A = Na/\varphi$, where $a$ represents the area of a single particle, and we set $a$ to unity. 
As an illustrative example, in Fig.~\ref{fig_triangle}(a), we present a two-dimensional non-equilateral triangle system, with side lengths arranged in a 4:5:6 ratio. 
To account for periodicity, we employed a square boundary with periodic conditions, ensuring that any portion of a particle extending beyond the boundary is seamlessly mapped to the opposite side of the box. As depicted in the figure, the outer portion of a triangle at the upper boundary (indicated by dashed lines) is mapped to the lower boundary, and likewise, the outer portion of a triangle at the left boundary (also indicated by dashed lines) is mapped to the right boundary. 
Within the interval of $1 \le i \le N$, we computed $P_{i}$ for specific $i$ values. To achieve this, we generated configurations of $(i-1)$ particles with random positions and orientations (as depicted in Fig.~\ref{fig_triangle}(a)). Subsequently, we ran Monte Carlo (MC) simulations, employing a Widom insertion procedure \cite{Widom1963, Frenkel1996}.  Each MC step involved particle translation and particle rotation. 
The Widom insertion procedure was carried out to attempt the insertion of an additional triangle with a random position and orientation.
The process was categorized as either a success or a failure, contingent upon whether it intersects with existing triangles. In the event that an attempted triangle insertion resulted in overlap with one or more existing triangles, as illustrated by the red triangle in Fig.~\ref{fig_triangle}(b), it was classified as a failure. Conversely, if the insertion did not intersect with any prior triangles, exemplified by the green triangle in Fig.~\ref{fig_triangle}(b), it was deemed a success. 
We performed $10^{3}$ Widom insertions every 5 MC steps, resulting in a total of $8\times10^8$ attempted insertions. The probability $P_{i}$ was computed as the ratio of successful insertions to the total attempted insertions. 

In Fig.~\ref{fig_triangle}(c), we present the results for a system comprising $N=5000$ triangles and packing densities ranging from 0.1 to 0.5. We measured $P_{i}$ at intervals of $i=200$ and plotted $\ln P_{i}$ in the figure, depicted as square symbols. 
We observed that, for the same $i$ value, a higher packing density $\varphi$ corresponds to a smaller $\ln P_{i}$ value. This relationship arises because a higher packing density $\varphi$ results in a smaller box area, making it more difficult to insert triangles.
Additionally, we noticed that as $i$ increases, $\ln P_{i}$ gradually decreases from zero, and this decreasing trend is more pronounced at higher packing densities. This phenomenon is expected, as with a fixed box area, it becomes progressively more challenging to insert an additional triangle when there are already more triangles present, and the situation becomes even more challenging when the box area is smaller.
To achieve $\ln P_{i}$ values across the entire range from 1 to 5000, we conducted a fourth-order polynomial fit to the available data points, resulting in the smooth curves illustrated in the figure. The coefficients of the polynomial fitting curves can be found in Section I of the Supplementary Material. To calculate $P_{N+1}$, we conducted a similar measurement but increased the total number of insertions to $8\times10^9$ to improve precision. 
In Fig.~\ref{fig_triangle}(d), we plot the left side of Eq.(\ref{equ_relation}), represented as $\ln (\prod_{i=1}^{N} P_{N+1}/P_{i} )/N$, as a function of $1/N$, with square symbols denoting data points for various values of $N$. We observed that as the value of $N$ increases, resulting in a smaller $1/N$, the $\ln (\prod_{i=1}^{N} P_{N+1}/P_{i} )/N$ also increases.
Furthermore, the data presented in the figure demonstrates a clear linear trend. Consequently, we conducted a linear fitting analysis, represented by the straight lines in the figure. The fitting coefficients can be found in Table \ref{table_linear_coefficients}. It's important to note that the intercept of the linear fit represents the value of the function $\ln (\prod_{i=1}^{N} P_{N+1}/P_{i} )/N$ in the thermodynamic limit.

To determine the value of $\tilde{s}(0+)$ on the right-hand side of Eq.~(\ref{equ_relation}), we initially computed the scale distribution function $\tilde{s}(x)$. For each pair of particles, denoted as $i$ and $j$, there exists a minimum value of $x_{ij}$ that leads to an overlap between particle $i$ and particle $j$ when their separation is scaled as $(1-x_{ij})$ times their original distance, while retaining their orientations unchanged. 
For each individual particle $i$, $\tilde{s}(x)$ identifies the minimum value $x_{i}$ from all the pairs formed by particle $i$ and another particle. Essentially, $x_{i}$ represents the scaling factor for particle $i$ concerning its nearest neighbor. Each particle contributes one count to the histogram of $\tilde{s}(x)$. 
To obtain $\tilde{s}(x)$, we conducted a dedicated MC simulation, where we computed $\tilde{s}(x)$ at intervals of 5 steps and performed a total of $4\times10^{10}/N$ steps. This choice was made because the scale distribution function characterizes the probability distribution of particle overlap with their nearest neighbors, and each particle contributes one count to the statistics of $\tilde{s}(x)$, as mentioned above. Systems with fewer particles necessitated additional simulation steps to achieve comparable accuracy. 
We computed $\tilde{s}(x)$ as a function of the scale factor $x$ within 100 bins, using a fixed $x_\mathrm{max}=0.05$.

In Fig.~\ref{fig_sdf}(a), we present the histogram of $\tilde{s}(x)$ for the two-dimensional non-equilateral triangle system with $N = 5000$, denoted as circles in the figure. 
As observed in the figure, across all packing densities, $\tilde{s}(x)$ exhibits a smooth relationship with $x$. In addition, as packing densities increase, $\tilde{s}(x)$ displays larger values, indicating a greater number of counts falling within this range overall. Furthermore, as packing densities increase, it becomes more pronounced that $\tilde{s}(x)$ exhibits a decreasing trend with $x$. This indicates that the separations between particles and their nearest neighbors are diminishing.
We then applied a fourth-order polynomial fit, $\tilde{s}(x)= B_0 + B_1x + B_2x^{2} + B_3x^{3} + B_4x^{4}$, to the data points, as represented by the curves in the figure. The coefficients of this fitting are provided in Section I of the Supplementary Material. The constant term, $B_{0}$ of this fitting, serves as the extrapolated value for $\tilde{s}(0+)$.
Figure \ref{fig_sdf}(b) displays $\tilde{s}(0+)$ as a function of $1/N$ for various values of $N$. Similar to the left side of the equation, the data points again exhibit a linear trend. However, in contrast to Fig.~\ref{fig_triangle}(d), where the data points displayed a negative slope, here they show a positive slope.
Furthermore, for the same packing fraction, as $N$ attains larger values, the data points within the figure and those from Fig.~\ref{fig_triangle}(d) exhibit a convergence towards closely matching values. We conducted a linear fitting analysis on these data points and provide the fitting coefficients in Table \ref{table_linear_coefficients}.

\begin{table*}
\caption{Linear fitting coefficients for $\ln (\prod_{i=1}^{N} P_{N+1}/P_{i} )/N$ and $-\tilde{s}(0+)/2d$ as a function of $1/N$ in the non-equilateral triangle system with a side ratio of 4:5:6. Numbers in brackets are the standard error.}
\begin{center}
\begin{tabular}{ p{1.5cm} p{2.5cm} p{3.5cm} p{2.5cm}}
 \hline \hline
$\varphi$&             & $\ln (\prod_{i=1}^{N} P_{N+1}/P_{i} )/N$ & \hspace{0em} $-\tilde{s}(0+)/2d$   \\
\hline
 0.1  & slope        &  \hspace{1em} -0.3794(40)                   &  \hspace{0.0em} 0.313(18)\\
      & intercept    &  \hspace{1em} \textbf{-0.348643(7)}         &  \textbf{-0.34866(6)}\\
 0.2  & slope        &  \hspace{1em} -1.0977(91)                   &  \hspace{0.0em} 0.581(26)\\
      & intercept    &  \hspace{1em} \textbf{-0.889678(18)}        &  \textbf{-0.88969(8)}\\
 0.3  & slope        &  \hspace{1em} -2.484(21)                    &  \hspace{0.0em} 0.983(38)\\
      & intercept    &  \hspace{1em} \textbf{-1.744382(37)}        &  \textbf{-1.74454(13)}\\
 0.4  & slope        &  \hspace{1em} -4.923(45)                    &  \hspace{0.0em} 1.069(54)\\
      & intercept    &  \hspace{1em} \textbf{-3.086051(96)}        &  \textbf{-3.08585(16)}\\
 0.5  & slope        &  \hspace{1em} -8.84(14)                     &  \hspace{0.0em} 1.23(11)\\
      & intercept    &  \hspace{1em} \textbf{-5.09853(33)}         &  \textbf{-5.09841(35)}\\
\hline
\end{tabular}
\end{center}
\label{table_linear_coefficients}
\end{table*}

\begin{table*}
\renewcommand{\arraystretch}{1.4}
\caption{Values of both sides of Eq.~(\ref{equ_relation}) in the thermodynamic limit for various hard particle systems. Numbers in brackets indicate standard errors. }
\begin{center}
\begin{tabular}{p{0.6cm}p{2.7cm}p{2.25cm}p{2.25cm}p{2.25cm}p{2.25cm}p{2.3cm}p{2.2cm}}
 \hline \hline
 &  & \hspace{0.0em} Segments & \hspace{0.5em} Disks   & \hspace{0.0em}Equilateral \hspace{0.0em}triangles & \hspace{0.0em} Squares  & Rectangles (side ratio 1:2) & \hspace{0.5em} Spheres\\ 
 $\varphi$& & $\hspace{1.0em} d=1$& $\hspace{1.0em} d=2$& $\hspace{1.0em} d=2$& $\hspace{1.0em} d=2$& $\hspace{1.0em} d=2$&$\hspace{1.0em} d=3$\\
 \hline
 0.1  &  $\frac{1}{N} \ln (\prod_{i=1}^{N} \frac{P_{N+1}}{P_{i}} ) $    & -0.222225(4)   &-0.236140(5)           &-0.327687(6)         &-0.274004(6)       &-0.295275(6)    &-0.521653(7)\\
      & $-\tilde{s}(0+)/2d$                                             & -0.222201(41)  &-0.236174(34)          &-0.327747(44)        &-0.273985(45)      &-0.295337(44)   &-0.521777(90)\\
\hline
 0.2  &  $\frac{1}{N} \ln (\prod_{i=1}^{N} \frac{P_{N+1}}{P_{i}} ) $    & -0.499987(10)  &-0.570367(10)          &-0.831744(13)        &-0.678307(11)      &-0.735896(14)   &-1.409420(23)\\
      & $-\tilde{s}(0+)/2d$                                             & -0.500031(54)  &-0.570341(67)          &-0.831703(81)        &-0.678227(69)      &-0.735879(61)   &-1.40906(19)\\
\hline
 0.3  &  $\frac{1}{N} \ln (\prod_{i=1}^{N} \frac{P_{N+1}}{P_{i}} ) $    & -0.857140(18)  &-1.063262(17)          &-1.628381(31)        &-1.300832(20)      &-1.418776(23)   &-2.984147(70)\\
      & $-\tilde{s}(0+)/2d$                                             & -0.857191(68)  &-1.063167(89)          &-1.62839(11)         &-1.300849(96)      &-1.41881(14)    &-2.98458(49)\\
\hline
 0.4  &  $\frac{1}{N} \ln (\prod_{i=1}^{N} \frac{P_{N+1}}{P_{i}} ) $    & -1.333298(35)  &-1.829281(33)          &-2.897497(82)        &-2.306821(42)      &-2.520731(58)   &-5.94685(51)\\
      & $-\tilde{s}(0+)/2d$                                             & -1.33337(9)    &-1.82930(13)           &-2.89740(17)         &-2.30681(14)       &-2.52075(14)    &-5.9449(14)\\
\hline
 0.5  &  $\frac{1}{N} \ln (\prod_{i=1}^{N} \frac{P_{N+1}}{P_{i}} ) $    & -1.99987(11)   &-3.106517(75)          &-4.84807(27)         &-4.01391(18)       &-4.36268(21)    &-12.058(37)\\
      & $-\tilde{s}(0+)/2d$                                             & -2.00019(25)   &-3.10620(23)           &-4.84777(20)         &-4.01397(23)       &-4.36233(22)    &-12.0174(69)\\
\hline
\end{tabular}
\end{center}
\label{table_other_shapes}
\end{table*}

As shown in Table \ref{table_linear_coefficients}, the intercepts of both linear fits (highlighted in bold) yield values that are remarkably close. This observation suggests that the two sides of Eq.~(\ref{equ_relation}) exhibit very similar values in the thermodynamic limit.
We then proceeded to assess the validity of Eq.~(\ref{equ_relation}) with respect to other particle shapes and dimensions. In two-dimensional systems, we investigated disks, equilateral triangles, squares, and rectangles with a side ratio of 1:2. Additionally, we explored one-dimensional line segments and three-dimensional spheres. For all these cases, we calculated both sides of Eq.~(\ref{equ_relation}) and provided the values in the thermodynamic limit in Table \ref{table_other_shapes}.
As seen from the table, despite differences in dimensionality and shape characteristics, both sides of Eq.~(\ref{equ_relation}) consistently exhibited close alignment for all the cases examined.

The discovery of Eq.~(\ref{equ_relation}) offers several advantages. On one hand, the values on both sides of the relation provide insights into the system's characteristics.
Firstly, both sides of the equation exhibit a decreasing trend as the system's density increases, allowing us to assess the degree of packing density within a system \cite{Labik1994, Maciel2018}. 
Secondly, the values on both sides are affected by particle shape, providing insights into the extent to which particles deviate from a ``round" shape. In two-dimensional cases, as demonstrated in Table \ref{table_other_shapes}, it becomes evident that disks exhibit larger values compared to squares at the same packing density, squares have larger values compared to rectangles with a side ratio of 1:2, followed by equilateral triangles. Notably, non-equilateral triangles with a side ratio of 4:5:6 display the smallest values. This observation aligns with the intuitive notion that inserting a disk is generally easier than a particle with a substantial departure from a round shape, particularly at high packing densities. Moreover, the equality of the two sides implies that in systems with non-round particles, we can expect the separation between a particle and its nearest neighbor to be smaller under the same packing density.
On the other hand, the equality of the two sides of the relation holds practical utility.
For instance, determining $\tilde{s}(0+)$ from $\tilde{s}(x)$ can be inherently noisy due to sampling characteristics, influenced by factors such as the choice of $x_\mathrm{max}$ and the number of bins (\cite{hoomdblue, Michael2013_pressure, Anderson2016}). Therefore, in cases where the left side of the equation can be accurately computed, it provides valuable guidance on selecting optimal sampling parameters.
Additionally, if we possess knowledge of $-\tilde{s}(0+)$ and all the $P_i$ values (where $1 \le i \le N$), the relation allows us to estimate $P_{N+1}$ and, thus, determine how many attempted insertions are needed for the successful placement of the $(N+1)$th particle.

At the last part of this letter, we provide an intuitive picture for understanding this relation. Though both quantities in the relation, i.e., the successful probability of inserting an additional particle and the scale distribution function, depend on particle shape and are thus geometrically induced, the equality of the relation is rooted in thermodynamics.
The term $\ln P_{N+1}$ on the left-hand side of the equation can be directly connected to the excess chemical potential $\mu_{\mathrm{ex}}$ of an $N$-particle system. This excess chemical potential represents the difference in chemical potential between the hard particle system and the corresponding ideal gas system under the same conditions, and the relationship is established previously \cite{Widom1963, Frenkel1996}:
\begin{equation}
\mu_{\mathrm{ex}} = -k_\mathrm{B}T  \ln P_{N+1}.
\label{chemical_potential} 
\end{equation}
Here the subscript ``ex" indicates the contribution beyond that of the ideal gas, $k_\mathrm{B}$ represents Boltzmann’s constant, and $T$ denotes temperature. Additionally, we established that the product of $\ln P_{i}$ on the left-hand side of the equation can be associated with the excess entropy of the system (see Section II of the Supplementary Material for details): 
\begin{equation}
S_{\mathrm{ex}} =  k_\mathrm{B} \ln (\prod_{i=1}^{N} P_{i}).
\label{s_entropy} 
\end{equation}
On the other hand, the scale distribution function $\tilde{s}(0+)$ on the right-hand side of the equation can be connected to the pressure of a convex hard particle system \cite{Eppenga1984, Brumby2011, Anderson2016}:
\begin{equation}
p = \frac{N k_{B}T }{V} (1+ \frac{\tilde{s}(0+) }{2d} ).
\label{pressure} 
\end{equation}
We consider the thermodynamic relation $\mu N = U - TS +PV$, where $\mu$ represents the chemical potential, $S$ is the entropy, and $p$ is the system's pressure. When using the ideal gas as a reference system, this relation can be expressed as:
\begin{equation}
\mu_{\mathrm{ex}} N = - TS_{\mathrm{ex}} +(PV-N k_{B}T )
\label{combination}
\end{equation}
Here we utilize the fact that both the ideal gas and hard particle systems possess zero potential energy and the same kinetic energy at the same temperature. We also apply the relation $pV=Nk_{\mathrm{B}}T$ for an ideal gas. Subsequently, by substituting Eqs.~(\ref{chemical_potential}), (\ref{s_entropy}), (\ref{pressure}) into Eq.~(\ref{combination}), we arrive at the originally stated Eq.~(\ref{equ_relation}). 
While the relation primarily involves only two quantities, its foundation in thermodynamics encompasses three thermodynamic variables. However, this is not a typical thermodynamic relation connecting the chemical potential, entropy, and pressure of a hard particle system, but rather, it pertains to their excess values relative to an ideal gas system.

In conclusion, we have uncovered a concise relation between two significant quantities within hard particle systems: the probability of successfully inserting an additional particle at random and the scale distribution function. We anticipate that this relation holds universal applicability for convex hard particles. To assess the validity of it, we conducted tests across a range of particle shapes, encompassing one-dimensional line segments, two-dimensional disks, equilateral and non-equilateral triangles, squares, rectangles, and three-dimensional spheres. Remarkably, we observed close agreement between the two sides of the equation in all examined cases. Moreover, we have demonstrated that this relationship can be derived from the fundamental thermodynamic relation that establishes connections between entropy, pressure, and chemical potential. Given that both of these quantities essentially reflect the packing density of a hard particle system and are intimately related to particle shape, our study reveals a geometrically rooted relation that underlies fundamental thermodynamic principles. This contribution sheds light on the intricate interplay between geometry and thermodynamics within hard particle systems.

\acknowledgments  
 This work was supported by the National Natural Science Foundation of China (Grant No.~12274330) and the Knowledge Innovation Program of Wuhan-Shuguang (Grant No.~2022010801020125).

\bibliography{geometry}

\begin{thebibliography}{38}%
\makeatletter
\providecommand \@ifxundefined [1]{%
 \@ifx{#1\undefined}
}%
\providecommand \@ifnum [1]{%
 \ifnum #1\expandafter \@firstoftwo
 \else \expandafter \@secondoftwo
 \fi
}%
\providecommand \@ifx [1]{%
 \ifx #1\expandafter \@firstoftwo
 \else \expandafter \@secondoftwo
 \fi
}%
\providecommand \natexlab [1]{#1}%
\providecommand \enquote  [1]{``#1''}%
\providecommand \bibnamefont  [1]{#1}%
\providecommand \bibfnamefont [1]{#1}%
\providecommand \citenamefont [1]{#1}%
\providecommand \href@noop [0]{\@secondoftwo}%
\providecommand \href [0]{\begingroup \@sanitize@url \@href}%
\providecommand \@href[1]{\@@startlink{#1}\@@href}%
\providecommand \@@href[1]{\endgroup#1\@@endlink}%
\providecommand \@sanitize@url [0]{\catcode `\\12\catcode `\$12\catcode `\&12\catcode `\#12\catcode `\^12\catcode `\_12\catcode `\%12\relax}%
\providecommand \@@startlink[1]{}%
\providecommand \@@endlink[0]{}%
\providecommand \url  [0]{\begingroup\@sanitize@url \@url }%
\providecommand \@url [1]{\endgroup\@href {#1}{\urlprefix }}%
\providecommand \urlprefix  [0]{URL }%
\providecommand \Eprint [0]{\href }%
\providecommand \doibase [0]{https://doi.org/}%
\providecommand \selectlanguage [0]{\@gobble}%
\providecommand \bibinfo  [0]{\@secondoftwo}%
\providecommand \bibfield  [0]{\@secondoftwo}%
\providecommand \translation [1]{[#1]}%
\providecommand \BibitemOpen [0]{}%
\providecommand \bibitemStop [0]{}%
\providecommand \bibitemNoStop [0]{.\EOS\space}%
\providecommand \EOS [0]{\spacefactor3000\relax}%
\providecommand \BibitemShut  [1]{\csname bibitem#1\endcsname}%
\let\auto@bib@innerbib\@empty
\bibitem [{\citenamefont {Manoharan}(2015)}]{Manoharan2015}%
  \BibitemOpen
  \bibfield  {author} {\bibinfo {author} {\bibfnamefont {V.~N.}\ \bibnamefont {Manoharan}},\ }\href@noop {} {\bibfield  {journal} {\bibinfo  {journal} {Science}\ }\textbf {\bibinfo {volume} {349}},\ \bibinfo {pages} {1253751} (\bibinfo {year} {2015})}\BibitemShut {NoStop}%
\bibitem [{\citenamefont {Torquato}\ and\ \citenamefont {Jiao}(2009)}]{Torquato2009}%
  \BibitemOpen
  \bibfield  {author} {\bibinfo {author} {\bibfnamefont {S.}~\bibnamefont {Torquato}}\ and\ \bibinfo {author} {\bibfnamefont {Y.}~\bibnamefont {Jiao}},\ }\href@noop {} {\bibfield  {journal} {\bibinfo  {journal} {Nature}\ }\textbf {\bibinfo {volume} {460}},\ \bibinfo {pages} {876} (\bibinfo {year} {2009})}\BibitemShut {NoStop}%
\bibitem [{\citenamefont {Haji-Akbari}\ \emph {et~al.}(2009)\citenamefont {Haji-Akbari}, \citenamefont {Engel}, \citenamefont {Keys}, \citenamefont {Zheng}, \citenamefont {Petschek}, \citenamefont {Palffy-Muhoray},\ and\ \citenamefont {Glotzer}}]{Haji-Akbari2009}%
  \BibitemOpen
  \bibfield  {author} {\bibinfo {author} {\bibfnamefont {A.}~\bibnamefont {Haji-Akbari}}, \bibinfo {author} {\bibfnamefont {M.}~\bibnamefont {Engel}}, \bibinfo {author} {\bibfnamefont {A.~S.}\ \bibnamefont {Keys}}, \bibinfo {author} {\bibfnamefont {X.}~\bibnamefont {Zheng}}, \bibinfo {author} {\bibfnamefont {R.~G.}\ \bibnamefont {Petschek}}, \bibinfo {author} {\bibfnamefont {P.}~\bibnamefont {Palffy-Muhoray}},\ and\ \bibinfo {author} {\bibfnamefont {S.~C.}\ \bibnamefont {Glotzer}},\ }\href@noop {} {\bibfield  {journal} {\bibinfo  {journal} {Nature}\ }\textbf {\bibinfo {volume} {462}},\ \bibinfo {pages} {773} (\bibinfo {year} {2009})}\BibitemShut {NoStop}%
\bibitem [{\citenamefont {Marechal}\ and\ \citenamefont {Dijkstra}(2010)}]{Marechal2010}%
  \BibitemOpen
  \bibfield  {author} {\bibinfo {author} {\bibfnamefont {M.}~\bibnamefont {Marechal}}\ and\ \bibinfo {author} {\bibfnamefont {M.}~\bibnamefont {Dijkstra}},\ }\href@noop {} {\bibfield  {journal} {\bibinfo  {journal} {Phys. Rev. E}\ }\textbf {\bibinfo {volume} {82}},\ \bibinfo {pages} {031405} (\bibinfo {year} {2010})}\BibitemShut {NoStop}%
\bibitem [{\citenamefont {Gang}\ and\ \citenamefont {Zhang}(2011)}]{Gang2011}%
  \BibitemOpen
  \bibfield  {author} {\bibinfo {author} {\bibfnamefont {O.}~\bibnamefont {Gang}}\ and\ \bibinfo {author} {\bibfnamefont {Y.}~\bibnamefont {Zhang}},\ }\href@noop {} {\bibfield  {journal} {\bibinfo  {journal} {ACS Nano}\ }\textbf {\bibinfo {volume} {5}},\ \bibinfo {pages} {8459} (\bibinfo {year} {2011})}\BibitemShut {NoStop}%
\bibitem [{\citenamefont {Agarwal}\ and\ \citenamefont {Escobedo}(2011)}]{Agarwal2011}%
  \BibitemOpen
  \bibfield  {author} {\bibinfo {author} {\bibfnamefont {U.}~\bibnamefont {Agarwal}}\ and\ \bibinfo {author} {\bibfnamefont {F.~A.}\ \bibnamefont {Escobedo}},\ }\href@noop {} {\bibfield  {journal} {\bibinfo  {journal} {Nat. Mater.}\ }\textbf {\bibinfo {volume} {10}},\ \bibinfo {pages} {230} (\bibinfo {year} {2011})}\BibitemShut {NoStop}%
\bibitem [{\citenamefont {Damasceno}\ \emph {et~al.}(2012)\citenamefont {Damasceno}, \citenamefont {Engel},\ and\ \citenamefont {Glotzer}}]{Damasceno2012}%
  \BibitemOpen
  \bibfield  {author} {\bibinfo {author} {\bibfnamefont {P.~F.}\ \bibnamefont {Damasceno}}, \bibinfo {author} {\bibfnamefont {M.}~\bibnamefont {Engel}},\ and\ \bibinfo {author} {\bibfnamefont {S.~C.}\ \bibnamefont {Glotzer}},\ }\href@noop {} {\bibfield  {journal} {\bibinfo  {journal} {Science}\ }\textbf {\bibinfo {volume} {337}},\ \bibinfo {pages} {453} (\bibinfo {year} {2012})}\BibitemShut {NoStop}%
\bibitem [{\citenamefont {Ni}\ \emph {et~al.}(2012)\citenamefont {Ni}, \citenamefont {Gantapara}, \citenamefont {de~Graaf}, \citenamefont {van Roij},\ and\ \citenamefont {Dijkstra}}]{Ni2012}%
  \BibitemOpen
  \bibfield  {author} {\bibinfo {author} {\bibfnamefont {R.}~\bibnamefont {Ni}}, \bibinfo {author} {\bibfnamefont {A.~P.}\ \bibnamefont {Gantapara}}, \bibinfo {author} {\bibfnamefont {J.}~\bibnamefont {de~Graaf}}, \bibinfo {author} {\bibfnamefont {R.}~\bibnamefont {van Roij}},\ and\ \bibinfo {author} {\bibfnamefont {M.}~\bibnamefont {Dijkstra}},\ }\href@noop {} {\bibfield  {journal} {\bibinfo  {journal} {Soft Matter}\ }\textbf {\bibinfo {volume} {8}},\ \bibinfo {pages} {8826} (\bibinfo {year} {2012})}\BibitemShut {NoStop}%
\bibitem [{\citenamefont {Smallenburg}\ \emph {et~al.}(2012)\citenamefont {Smallenburg}, \citenamefont {Filion}, \citenamefont {Marechal},\ and\ \citenamefont {Dijkstra}}]{Smallenburg2012}%
  \BibitemOpen
  \bibfield  {author} {\bibinfo {author} {\bibfnamefont {F.}~\bibnamefont {Smallenburg}}, \bibinfo {author} {\bibfnamefont {L.}~\bibnamefont {Filion}}, \bibinfo {author} {\bibfnamefont {M.}~\bibnamefont {Marechal}},\ and\ \bibinfo {author} {\bibfnamefont {M.}~\bibnamefont {Dijkstra}},\ }\href@noop {} {\bibfield  {journal} {\bibinfo  {journal} {Proc. Natl. Acad. Sci.}\ }\textbf {\bibinfo {volume} {109}},\ \bibinfo {pages} {17886} (\bibinfo {year} {2012})}\BibitemShut {NoStop}%
\bibitem [{\citenamefont {Avendaño}\ and\ \citenamefont {Escobedo}(2012)}]{Avendano2012}%
  \BibitemOpen
  \bibfield  {author} {\bibinfo {author} {\bibfnamefont {C.}~\bibnamefont {Avendaño}}\ and\ \bibinfo {author} {\bibfnamefont {F.~A.}\ \bibnamefont {Escobedo}},\ }\href@noop {} {\bibfield  {journal} {\bibinfo  {journal} {Soft Matter}\ }\textbf {\bibinfo {volume} {8}},\ \bibinfo {pages} {4675} (\bibinfo {year} {2012})}\BibitemShut {NoStop}%
\bibitem [{\citenamefont {Gantapara}\ \emph {et~al.}(2013)\citenamefont {Gantapara}, \citenamefont {de~Graaf}, \citenamefont {van Roij},\ and\ \citenamefont {Dijkstra}}]{Gantapara2013}%
  \BibitemOpen
  \bibfield  {author} {\bibinfo {author} {\bibfnamefont {A.~P.}\ \bibnamefont {Gantapara}}, \bibinfo {author} {\bibfnamefont {J.}~\bibnamefont {de~Graaf}}, \bibinfo {author} {\bibfnamefont {R.}~\bibnamefont {van Roij}},\ and\ \bibinfo {author} {\bibfnamefont {M.}~\bibnamefont {Dijkstra}},\ }\href@noop {} {\bibfield  {journal} {\bibinfo  {journal} {Phys. Rev. Lett.}\ }\textbf {\bibinfo {volume} {111}},\ \bibinfo {pages} {015501} (\bibinfo {year} {2013})}\BibitemShut {NoStop}%
\bibitem [{\citenamefont {Bernard}\ and\ \citenamefont {Krauth}(2011)}]{Bernard2011_melting}%
  \BibitemOpen
  \bibfield  {author} {\bibinfo {author} {\bibfnamefont {E.~P.}\ \bibnamefont {Bernard}}\ and\ \bibinfo {author} {\bibfnamefont {W.}~\bibnamefont {Krauth}},\ }\href@noop {} {\bibfield  {journal} {\bibinfo  {journal} {Phys. Rev. Lett.}\ }\textbf {\bibinfo {volume} {107}},\ \bibinfo {pages} {155704} (\bibinfo {year} {2011})}\BibitemShut {NoStop}%
\bibitem [{\citenamefont {Anderson}\ \emph {et~al.}(2017)\citenamefont {Anderson}, \citenamefont {Antonaglia}, \citenamefont {Millan}, \citenamefont {Engel},\ and\ \citenamefont {Glotzer}}]{Anderson2017}%
  \BibitemOpen
  \bibfield  {author} {\bibinfo {author} {\bibfnamefont {J.~A.}\ \bibnamefont {Anderson}}, \bibinfo {author} {\bibfnamefont {J.}~\bibnamefont {Antonaglia}}, \bibinfo {author} {\bibfnamefont {J.~A.}\ \bibnamefont {Millan}}, \bibinfo {author} {\bibfnamefont {M.}~\bibnamefont {Engel}},\ and\ \bibinfo {author} {\bibfnamefont {S.~C.}\ \bibnamefont {Glotzer}},\ }\href@noop {} {\bibfield  {journal} {\bibinfo  {journal} {Phys. Rev. X}\ }\textbf {\bibinfo {volume} {7}},\ \bibinfo {pages} {021001} (\bibinfo {year} {2017})}\BibitemShut {NoStop}%
\bibitem [{\citenamefont {Lei}\ \emph {et~al.}(2018)\citenamefont {Lei}, \citenamefont {Ni},\ and\ \citenamefont {Ma}}]{Lei2018_helix}%
  \BibitemOpen
  \bibfield  {author} {\bibinfo {author} {\bibfnamefont {Q.-l.}\ \bibnamefont {Lei}}, \bibinfo {author} {\bibfnamefont {R.}~\bibnamefont {Ni}},\ and\ \bibinfo {author} {\bibfnamefont {Y.-q.}\ \bibnamefont {Ma}},\ }\href@noop {} {\bibfield  {journal} {\bibinfo  {journal} {ACS Nano}\ }\textbf {\bibinfo {volume} {12}},\ \bibinfo {pages} {6860} (\bibinfo {year} {2018})}\BibitemShut {NoStop}%
\bibitem [{\citenamefont {Klotsa}\ \emph {et~al.}(2018)\citenamefont {Klotsa}, \citenamefont {Chen}, \citenamefont {Engel},\ and\ \citenamefont {Glotzer}}]{Klotsa2018}%
  \BibitemOpen
  \bibfield  {author} {\bibinfo {author} {\bibfnamefont {D.}~\bibnamefont {Klotsa}}, \bibinfo {author} {\bibfnamefont {E.~R.}\ \bibnamefont {Chen}}, \bibinfo {author} {\bibfnamefont {M.}~\bibnamefont {Engel}},\ and\ \bibinfo {author} {\bibfnamefont {S.~C.}\ \bibnamefont {Glotzer}},\ }\href@noop {} {\bibfield  {journal} {\bibinfo  {journal} {Soft Matter}\ }\textbf {\bibinfo {volume} {14}},\ \bibinfo {pages} {8692} (\bibinfo {year} {2018})}\BibitemShut {NoStop}%
\bibitem [{\citenamefont {Wan}\ \emph {et~al.}(2019)\citenamefont {Wan}, \citenamefont {Du}, \citenamefont {van Anders},\ and\ \citenamefont {Glotzer}}]{Wan2019}%
  \BibitemOpen
  \bibfield  {author} {\bibinfo {author} {\bibfnamefont {D.}~\bibnamefont {Wan}}, \bibinfo {author} {\bibfnamefont {C.~X.}\ \bibnamefont {Du}}, \bibinfo {author} {\bibfnamefont {G.}~\bibnamefont {van Anders}},\ and\ \bibinfo {author} {\bibfnamefont {S.~C.}\ \bibnamefont {Glotzer}},\ }\href@noop {} {\bibfield  {journal} {\bibinfo  {journal} {J. Phys. Chem. B}\ }\textbf {\bibinfo {volume} {123}},\ \bibinfo {pages} {9038} (\bibinfo {year} {2019})}\BibitemShut {NoStop}%
\bibitem [{\citenamefont {Wan}\ and\ \citenamefont {Glotzer}(2021)}]{Wan2021}%
  \BibitemOpen
  \bibfield  {author} {\bibinfo {author} {\bibfnamefont {D.}~\bibnamefont {Wan}}\ and\ \bibinfo {author} {\bibfnamefont {S.~C.}\ \bibnamefont {Glotzer}},\ }\href@noop {} {\bibfield  {journal} {\bibinfo  {journal} {Phys. Rev. Lett.}\ }\textbf {\bibinfo {volume} {126}},\ \bibinfo {pages} {208002} (\bibinfo {year} {2021})}\BibitemShut {NoStop}%
\bibitem [{\citenamefont {Carlsson}\ \emph {et~al.}(2012)\citenamefont {Carlsson}, \citenamefont {Gorham}, \citenamefont {Kahle},\ and\ \citenamefont {Mason}}]{Carlsson2012_topology}%
  \BibitemOpen
  \bibfield  {author} {\bibinfo {author} {\bibfnamefont {G.}~\bibnamefont {Carlsson}}, \bibinfo {author} {\bibfnamefont {J.}~\bibnamefont {Gorham}}, \bibinfo {author} {\bibfnamefont {M.}~\bibnamefont {Kahle}},\ and\ \bibinfo {author} {\bibfnamefont {J.}~\bibnamefont {Mason}},\ }\href@noop {} {\bibfield  {journal} {\bibinfo  {journal} {Phys. Rev. E}\ }\textbf {\bibinfo {volume} {85}},\ \bibinfo {pages} {011303} (\bibinfo {year} {2012})}\BibitemShut {NoStop}%
\bibitem [{\citenamefont {Blair}\ \emph {et~al.}(2012)\citenamefont {Blair}, \citenamefont {Santangelo},\ and\ \citenamefont {Machta}}]{Blair2012_squares}%
  \BibitemOpen
  \bibfield  {author} {\bibinfo {author} {\bibfnamefont {D.~W.}\ \bibnamefont {Blair}}, \bibinfo {author} {\bibfnamefont {C.}~\bibnamefont {Santangelo}},\ and\ \bibinfo {author} {\bibfnamefont {J.}~\bibnamefont {Machta}},\ }\href@noop {} {\bibfield  {journal} {\bibinfo  {journal} {J. Stat. Mech.}\ }\textbf {\bibinfo {volume} {2012}},\ \bibinfo {pages} {P01018} (\bibinfo {year} {2012})}\BibitemShut {NoStop}%
\bibitem [{\citenamefont {Teich}\ \emph {et~al.}(2016)\citenamefont {Teich}, \citenamefont {van Anders}, \citenamefont {Klotsa}, \citenamefont {Dshemuchadse},\ and\ \citenamefont {Glotzer}}]{Teich2016}%
  \BibitemOpen
  \bibfield  {author} {\bibinfo {author} {\bibfnamefont {E.~G.}\ \bibnamefont {Teich}}, \bibinfo {author} {\bibfnamefont {G.}~\bibnamefont {van Anders}}, \bibinfo {author} {\bibfnamefont {D.}~\bibnamefont {Klotsa}}, \bibinfo {author} {\bibfnamefont {J.}~\bibnamefont {Dshemuchadse}},\ and\ \bibinfo {author} {\bibfnamefont {S.~C.}\ \bibnamefont {Glotzer}},\ }\href@noop {} {\bibfield  {journal} {\bibinfo  {journal} {Proc. Natl. Acad. Sci.}\ }\textbf {\bibinfo {volume} {113}},\ \bibinfo {pages} {E669} (\bibinfo {year} {2016})}\BibitemShut {NoStop}%
\bibitem [{\citenamefont {Sitta}\ \emph {et~al.}(2018)\citenamefont {Sitta}, \citenamefont {Smallenburg}, \citenamefont {Wittkowski},\ and\ \citenamefont {L\"owen}}]{Sitta2018}%
  \BibitemOpen
  \bibfield  {author} {\bibinfo {author} {\bibfnamefont {C.~E.}\ \bibnamefont {Sitta}}, \bibinfo {author} {\bibfnamefont {F.}~\bibnamefont {Smallenburg}}, \bibinfo {author} {\bibfnamefont {R.}~\bibnamefont {Wittkowski}},\ and\ \bibinfo {author} {\bibfnamefont {H.}~\bibnamefont {L\"owen}},\ }\href@noop {} {\bibfield  {journal} {\bibinfo  {journal} {Phys. Chem. Chem. Phys.}\ }\textbf {\bibinfo {volume} {20}},\ \bibinfo {pages} {5285} (\bibinfo {year} {2018})}\BibitemShut {NoStop}%
\bibitem [{\citenamefont {Wan}\ and\ \citenamefont {Glotzer}(2018)}]{Wan2018}%
  \BibitemOpen
  \bibfield  {author} {\bibinfo {author} {\bibfnamefont {D.}~\bibnamefont {Wan}}\ and\ \bibinfo {author} {\bibfnamefont {S.~C.}\ \bibnamefont {Glotzer}},\ }\href@noop {} {\bibfield  {journal} {\bibinfo  {journal} {Soft Matter}\ }\textbf {\bibinfo {volume} {14}},\ \bibinfo {pages} {3012} (\bibinfo {year} {2018})}\BibitemShut {NoStop}%
\bibitem [{\citenamefont {Wan}(2022)}]{Wan2022}%
  \BibitemOpen
  \bibfield  {author} {\bibinfo {author} {\bibfnamefont {D.}~\bibnamefont {Wan}},\ }\href@noop {} {\bibfield  {journal} {\bibinfo  {journal} {Phys. Rev. E}\ }\textbf {\bibinfo {volume} {106}},\ \bibinfo {pages} {034609} (\bibinfo {year} {2022})}\BibitemShut {NoStop}%
\bibitem [{\citenamefont {Mughal}\ \emph {et~al.}(2012)\citenamefont {Mughal}, \citenamefont {Chan}, \citenamefont {Weaire},\ and\ \citenamefont {Hutzler}}]{Mughal2012_spheres}%
  \BibitemOpen
  \bibfield  {author} {\bibinfo {author} {\bibfnamefont {A.}~\bibnamefont {Mughal}}, \bibinfo {author} {\bibfnamefont {H.~K.}\ \bibnamefont {Chan}}, \bibinfo {author} {\bibfnamefont {D.}~\bibnamefont {Weaire}},\ and\ \bibinfo {author} {\bibfnamefont {S.}~\bibnamefont {Hutzler}},\ }\href@noop {} {\bibfield  {journal} {\bibinfo  {journal} {Phys. Rev. E}\ }\textbf {\bibinfo {volume} {85}},\ \bibinfo {pages} {051305} (\bibinfo {year} {2012})}\BibitemShut {NoStop}%
\bibitem [{\citenamefont {Jin}\ \emph {et~al.}(2020)\citenamefont {Jin}, \citenamefont {Chan},\ and\ \citenamefont {Zhong}}]{Jin2020_spheroids}%
  \BibitemOpen
  \bibfield  {author} {\bibinfo {author} {\bibfnamefont {W.}~\bibnamefont {Jin}}, \bibinfo {author} {\bibfnamefont {H.-K.}\ \bibnamefont {Chan}},\ and\ \bibinfo {author} {\bibfnamefont {Z.}~\bibnamefont {Zhong}},\ }\href@noop {} {\bibfield  {journal} {\bibinfo  {journal} {Phys. Rev. Lett.}\ }\textbf {\bibinfo {volume} {124}},\ \bibinfo {pages} {248002} (\bibinfo {year} {2020})}\BibitemShut {NoStop}%
\bibitem [{\citenamefont {Liu}\ \emph {et~al.}(2023)\citenamefont {Liu}, \citenamefont {Chan},\ and\ \citenamefont {Wan}}]{Wan2023_chiral}%
  \BibitemOpen
  \bibfield  {author} {\bibinfo {author} {\bibfnamefont {T.}~\bibnamefont {Liu}}, \bibinfo {author} {\bibfnamefont {H.-K.}\ \bibnamefont {Chan}},\ and\ \bibinfo {author} {\bibfnamefont {D.}~\bibnamefont {Wan}},\ }\href@noop {} {\bibfield  {journal} {\bibinfo  {journal} {Soft Matter}\ }\textbf {\bibinfo {volume} {19}},\ \bibinfo {pages} {7313} (\bibinfo {year} {2023})}\BibitemShut {NoStop}%
\bibitem [{\citenamefont {Mbah}\ \emph {et~al.}(2023)\citenamefont {Mbah}, \citenamefont {Wang}, \citenamefont {Englisch}, \citenamefont {Bommineni}, \citenamefont {Varela-Rosales}, \citenamefont {Spiecker}, \citenamefont {Vogel},\ and\ \citenamefont {Engel}}]{Mbah2023}%
  \BibitemOpen
  \bibfield  {author} {\bibinfo {author} {\bibfnamefont {C.~F.}\ \bibnamefont {Mbah}}, \bibinfo {author} {\bibfnamefont {J.}~\bibnamefont {Wang}}, \bibinfo {author} {\bibfnamefont {S.}~\bibnamefont {Englisch}}, \bibinfo {author} {\bibfnamefont {P.}~\bibnamefont {Bommineni}}, \bibinfo {author} {\bibfnamefont {N.~R.}\ \bibnamefont {Varela-Rosales}}, \bibinfo {author} {\bibfnamefont {E.}~\bibnamefont {Spiecker}}, \bibinfo {author} {\bibfnamefont {N.}~\bibnamefont {Vogel}},\ and\ \bibinfo {author} {\bibfnamefont {M.}~\bibnamefont {Engel}},\ }\href@noop {} {\bibfield  {journal} {\bibinfo  {journal} {Nat. Commun.}\ }\textbf {\bibinfo {volume} {14}},\ \bibinfo {pages} {5299} (\bibinfo {year} {2023})}\BibitemShut {NoStop}%
\bibitem [{\citenamefont {Widom}(1963)}]{Widom1963}%
  \BibitemOpen
  \bibfield  {author} {\bibinfo {author} {\bibfnamefont {B.}~\bibnamefont {Widom}},\ }\href@noop {} {\bibfield  {journal} {\bibinfo  {journal} {J. Chem. Phys.}\ }\textbf {\bibinfo {volume} {39}},\ \bibinfo {pages} {2808} (\bibinfo {year} {1963})}\BibitemShut {NoStop}%
\bibitem [{\citenamefont {Frenkel}\ and\ \citenamefont {Smit}(1996)}]{Frenkel1996}%
  \BibitemOpen
  \bibfield  {author} {\bibinfo {author} {\bibfnamefont {D.}~\bibnamefont {Frenkel}}\ and\ \bibinfo {author} {\bibfnamefont {B.}~\bibnamefont {Smit}},\ }\href@noop {} {\emph {\bibinfo {title} {Understanding Molecular Simulation : from Algorithms to Applications}}}\ (\bibinfo  {publisher} {Academic Press},\ \bibinfo {year} {1996})\BibitemShut {NoStop}%
\bibitem [{\citenamefont {Eppenga}\ and\ \citenamefont {Frenkel}(1984)}]{Eppenga1984}%
  \BibitemOpen
  \bibfield  {author} {\bibinfo {author} {\bibfnamefont {R.}~\bibnamefont {Eppenga}}\ and\ \bibinfo {author} {\bibfnamefont {D.}~\bibnamefont {Frenkel}},\ }\href@noop {} {\bibfield  {journal} {\bibinfo  {journal} {Mol. Phys.}\ }\textbf {\bibinfo {volume} {52}},\ \bibinfo {pages} {1303} (\bibinfo {year} {1984})}\BibitemShut {NoStop}%
\bibitem [{\citenamefont {Brumby}\ \emph {et~al.}(2011)\citenamefont {Brumby}, \citenamefont {Haslam}, \citenamefont {de~Miguel},\ and\ \citenamefont {Jackson}}]{Brumby2011}%
  \BibitemOpen
  \bibfield  {author} {\bibinfo {author} {\bibfnamefont {P.~E.}\ \bibnamefont {Brumby}}, \bibinfo {author} {\bibfnamefont {A.~J.}\ \bibnamefont {Haslam}}, \bibinfo {author} {\bibfnamefont {E.}~\bibnamefont {de~Miguel}},\ and\ \bibinfo {author} {\bibfnamefont {G.}~\bibnamefont {Jackson}},\ }\href@noop {} {\bibfield  {journal} {\bibinfo  {journal} {Mol. Phys.}\ }\textbf {\bibinfo {volume} {109}},\ \bibinfo {pages} {169} (\bibinfo {year} {2011})}\BibitemShut {NoStop}%
\bibitem [{\citenamefont {Anderson}\ \emph {et~al.}(2016)\citenamefont {Anderson}, \citenamefont {Irrgang},\ and\ \citenamefont {Glotzer}}]{Anderson2016}%
  \BibitemOpen
  \bibfield  {author} {\bibinfo {author} {\bibfnamefont {J.~A.}\ \bibnamefont {Anderson}}, \bibinfo {author} {\bibfnamefont {M.~E.}\ \bibnamefont {Irrgang}},\ and\ \bibinfo {author} {\bibfnamefont {S.~C.}\ \bibnamefont {Glotzer}},\ }\href@noop {} {\bibfield  {journal} {\bibinfo  {journal} {Comput. Phys. Commun.}\ }\textbf {\bibinfo {volume} {204}},\ \bibinfo {pages} {21} (\bibinfo {year} {2016})}\BibitemShut {NoStop}%
\bibitem [{\citenamefont {Glaser}\ \emph {et~al.}(2015)\citenamefont {Glaser}, \citenamefont {Nguyen}, \citenamefont {Anderson}, \citenamefont {Lui}, \citenamefont {Spiga}, \citenamefont {Millan}, \citenamefont {Morse},\ and\ \citenamefont {Glotzer}}]{Glaser2015}%
  \BibitemOpen
  \bibfield  {author} {\bibinfo {author} {\bibfnamefont {J.}~\bibnamefont {Glaser}}, \bibinfo {author} {\bibfnamefont {T.~D.}\ \bibnamefont {Nguyen}}, \bibinfo {author} {\bibfnamefont {J.~A.}\ \bibnamefont {Anderson}}, \bibinfo {author} {\bibfnamefont {P.}~\bibnamefont {Lui}}, \bibinfo {author} {\bibfnamefont {F.}~\bibnamefont {Spiga}}, \bibinfo {author} {\bibfnamefont {J.~A.}\ \bibnamefont {Millan}}, \bibinfo {author} {\bibfnamefont {D.~C.}\ \bibnamefont {Morse}},\ and\ \bibinfo {author} {\bibfnamefont {S.~C.}\ \bibnamefont {Glotzer}},\ }\href@noop {} {\bibfield  {journal} {\bibinfo  {journal} {Comput. Phys. Commun.}\ }\textbf {\bibinfo {volume} {192}},\ \bibinfo {pages} {97} (\bibinfo {year} {2015})}\BibitemShut {NoStop}%
\bibitem [{\citenamefont {Anderson}\ \emph {et~al.}(2020)\citenamefont {Anderson}, \citenamefont {Glaser},\ and\ \citenamefont {Glotzer}}]{Anderson2020}%
  \BibitemOpen
  \bibfield  {author} {\bibinfo {author} {\bibfnamefont {J.~A.}\ \bibnamefont {Anderson}}, \bibinfo {author} {\bibfnamefont {J.}~\bibnamefont {Glaser}},\ and\ \bibinfo {author} {\bibfnamefont {S.~C.}\ \bibnamefont {Glotzer}},\ }\href@noop {} {\bibfield  {journal} {\bibinfo  {journal} {Computati. Mater. Sci.}\ }\textbf {\bibinfo {volume} {173}},\ \bibinfo {pages} {109363} (\bibinfo {year} {2020})}\BibitemShut {NoStop}%
\bibitem [{\citenamefont {Lab\'ik}\ and\ \citenamefont {Smith}(1994)}]{Labik1994}%
  \BibitemOpen
  \bibfield  {author} {\bibinfo {author} {\bibfnamefont {S.}~\bibnamefont {Lab\'ik}}\ and\ \bibinfo {author} {\bibfnamefont {W.~R.}\ \bibnamefont {Smith}},\ }\href@noop {} {\bibfield  {journal} {\bibinfo  {journal} {Mol. Simul.}\ }\textbf {\bibinfo {volume} {12}},\ \bibinfo {pages} {23} (\bibinfo {year} {1994})}\BibitemShut {NoStop}%
\bibitem [{\citenamefont {Maciel}\ \emph {et~al.}(2018)\citenamefont {Maciel}, \citenamefont {Abreu},\ and\ \citenamefont {Tavares}}]{Maciel2018}%
  \BibitemOpen
  \bibfield  {author} {\bibinfo {author} {\bibfnamefont {J.~C. d. S.~L.}\ \bibnamefont {Maciel}}, \bibinfo {author} {\bibfnamefont {C.~R.~A.}\ \bibnamefont {Abreu}},\ and\ \bibinfo {author} {\bibfnamefont {F.~W.}\ \bibnamefont {Tavares}},\ }\href@noop {} {\bibfield  {journal} {\bibinfo  {journal} {Braz. J. Chem. Eng.}\ }\textbf {\bibinfo {volume} {35}},\ \bibinfo {pages} {277} (\bibinfo {year} {2018})}\BibitemShut {NoStop}%
\bibitem [{hoo()}]{hoomdblue}%
  \BibitemOpen
  \href@noop {} {\bibinfo {title} {Documentation of hoomd-blue}},\ \bibinfo {howpublished} {\url{http://glotzerlab.engin.umich.edu/hoomd-blue/}}\BibitemShut {NoStop}%
\bibitem [{\citenamefont {Engel}\ \emph {et~al.}(2013)\citenamefont {Engel}, \citenamefont {Anderson}, \citenamefont {Glotzer}, \citenamefont {Isobe}, \citenamefont {Bernard},\ and\ \citenamefont {Krauth}}]{Michael2013_pressure}%
  \BibitemOpen
  \bibfield  {author} {\bibinfo {author} {\bibfnamefont {M.}~\bibnamefont {Engel}}, \bibinfo {author} {\bibfnamefont {J.~A.}\ \bibnamefont {Anderson}}, \bibinfo {author} {\bibfnamefont {S.~C.}\ \bibnamefont {Glotzer}}, \bibinfo {author} {\bibfnamefont {M.}~\bibnamefont {Isobe}}, \bibinfo {author} {\bibfnamefont {E.~P.}\ \bibnamefont {Bernard}},\ and\ \bibinfo {author} {\bibfnamefont {W.}~\bibnamefont {Krauth}},\ }\href@noop {} {\bibfield  {journal} {\bibinfo  {journal} {Phys. Rev. E}\ }\textbf {\bibinfo {volume} {87}},\ \bibinfo {pages} {042134} (\bibinfo {year} {2013})}\BibitemShut {NoStop}%
\end{thebibliography}%

\newpage
\onecolumngrid

\titlespacing{\subsection}{0pt}{7.5ex plus 1ex minus .2ex}{2em}
\renewcommand{\theequation}{S\arabic{equation}}
\renewcommand{\thefigure}{S\arabic{figure}}
\renewcommand{\thetable}{S\arabic{table}}
\renewcommand{\thethm}{S\arabic{thm}}
\setcounter{equation}{0}
\setcounter{figure}{0}
\setcounter{table}{0}

\begin{center}
    {\bf \large Supplementary Material for:\\
A geometry-originated universal relation for arbitrary convex hard particles}
\end{center}

\setcounter{equation}{0}
\setcounter{figure}{0}
\setcounter{table}{0}
\setcounter{page}{1}
\makeatletter
\renewcommand{\theequation}{S\arabic{equation}}
\renewcommand{\thefigure}{S\arabic{figure}}
\renewcommand{\thetable}{S\arabic{table}}

\section{I. Polynomial Fitting and Fitting Coefficients}

We used a fourth-order polynomial fit to model the relationship between $\ln P_{i}$ and $i$ in Fig. 1(c) in the main text. The fitting coefficients are presented in Table \ref{fitting_coefficients}.

\begin{table*}[h]
\caption{Coefficients of the polynomial fits $\ln P_{i}= B_1*i + B_2*i^{2} + B_3*i^{3} + B_4*i^{4}$ applied to the data points in Fig.~1(c) of the main text.}
\begin{center}
\begin{tabular}{ p{1cm} p{3cm} p{3cm} p{3cm} p{3cm} p{3cm}}
 \hline \hline
$\hspace{0.2em} \phi$    &   &\hspace{2em} $B_1$        &\hspace{2em} $B_2$      &\hspace{2em} $B_3$        &\hspace{2em} $B_4$\\
\hline
0.1  & value        &-1.12162E-4    &-3.47084E-9    &-9.50479E-14   &-2.85907E-18\\
     &standard error&\hspace{0.0em} 3.08148E-8     &\hspace{0.0em} 3.27289E-11   &\hspace{0.0em} 1.08321E-14    &\hspace{0.0em} 1.12183E-18\\
\hline
0.2  & value        &-2.24329E-4    &-1.39152E-8   &-7.13333E-13   &-6.02486E-17\\
     &standard error&\hspace{0.0em} 4.59128E-8     &\hspace{0.0em} 4.87648E-11   &\hspace{0.0em} 1.61394E-14    &\hspace{0.0em} 1.67149E-18\\
\hline
0.3  & value        &-3.36100E-4    &-3.19673E-8   &-2.09771E-12   &-3.49443E-16\\
     &standard error&\hspace{0.0em} 1.35891E-7     &\hspace{0.0em} 1.44333E-10   &\hspace{0.0em} 4.77689E-14    &\hspace{0.0em} 4.94722E-18\\
\hline
0.4  & value        &-4.48722E-4    &-5.61405E-8   &-5.21117E-12   &-1.08139E-15\\
     &standard error&\hspace{0.0em} 4.04730E-7     &\hspace{0.0em} 4.29871E-10   &\hspace{0.0em} 1.42272E-13    &\hspace{0.0em} 1.47345E-17\\
\hline
0.5  & value        &-5.70600E-4    &-7.32400E-8   &-1.63872E-11   &-1.83927E-15\\
     &standard error&\hspace{0.0em} 2.81020E-6     &\hspace{0.0em} 2.98476E-9    &\hspace{0.0em} 9.87847E-13    &\hspace{0.0em} 1.02307E-16\\

\hline
\end{tabular}
\end{center}
\label{fitting_coefficients}
\end{table*}

We used a fourth-order polynomial fit to analyze the scale distribution function $\tilde{s}(x)$ in Fig.~2(a) in the main text. The fitting coefficients are presented in Table \ref{fitting_coefficients_sdf}.

\begin{table*}[h]
\caption{Coefficients of the polynomial fits $\tilde{s}(x)= B_0 + B_1*x + B_2*x^{2} + B_3*x^{3} + B_4*x^{4}$ for data points in Fig.~2(a) in the main text.}
\begin{center}
\begin{tabular}{p{1cm}p{2.5cm} p{2.5cm}p{2.5cm}p{2.5cm}p{2.5cm}p{2.5cm}}
 \hline \hline
$\hspace{0.2em} \phi$   &   &\hspace{1.5em} $B_0$         &\hspace{2em} $B_1$        &\hspace{2em} $B_2$      &\hspace{2em} $B_3$        &\hspace{2em} $B_4$\\
\hline
0.1  & value          &1.39459                 &\hspace{0.0em} 1.03020E-3   &\hspace{0.0em} 6.76766E-6                 & -1.00329E-7                                &-1.00329E-7\\
     &standard error  &4.25081E-4              &\hspace{0.0em} 5.89120E-5    &\hspace{0.0em} 2.39565E-6                &\hspace{0.0em} 3.59940E-8                    &\hspace{0.0em} 1.78561E-10\\
\hline
0.2  & value          &3.55739                 &-7.86356E-4                 &-1.48489E-6                               &-4.14963E-8                                 &-4.14963E-8\\
     &standard error  &6.43844E-4              &\hspace{0.0em} 8.92304E-5   &\hspace{0.0em} 3.62854E-6                 &\hspace{0.0em} 5.45179E-8                    &\hspace{0.0em} 2.70455E-10\\
\hline
0.3  & value          &6.97668                 &-0.0133185                  &-5.13390E-7                               &-5.89523E-8                                 &-5.89523E-8\\
     &standard error  &0.00103114              &\hspace{0.0em} 1.42905E-4   &\hspace{0.0em} 5.81122E-6                 &\hspace{0.0em} 8.73122E-8                    &\hspace{0.0em} 4.33142E-10\\
\hline
0.4  & value          &12.3404                 &-0.0562042                  &\hspace{0.0em} 6.77577E-5                 &\hspace{0.0em} 6.90502E-8                 &\hspace{0.0em} 6.90502E-8\\
     &standard error  &0.00129177              &\hspace{0.0em} 1.79026E-4   &\hspace{0.0em} 7.28005E-6                 &\hspace{0.0em} 1.09381E-7                    &\hspace{0.0em} 5.42623E-10\\
\hline
0.5  & value          &20.3928                 &-0.175207                   &\hspace{0.0em} 5.99135E-4                 &-1.01544E-6                                 &-1.01544E-6\\
     &standard error  &0.00181292              &\hspace{0.0em} 2.51252E-4   &\hspace{0.0em} 1.02171E-5                 &\hspace{0.0em} 1.53510E-7                    &\hspace{0.0em} 7.61540E-10\\

\hline
\end{tabular}
\end{center}
\label{fitting_coefficients_sdf}
\end{table*}

\section{II. Derivation of the excess entropy}
We consider a system of $N$ identical hard particles in a cubic box with length $L$ and volume $V_{\mathrm{box}}=L^{d}$, where $d$ is the dimension of the system. The centre-of-mass positions of the particles are given by the vector $\boldsymbol{r}$ and the orientations are given by $\boldsymbol{\omega}$. The momentum of the particles are $\boldsymbol{p}$, and the generalized momenta that corresponds to particle orientations are $\boldsymbol{L}$. The total internal energy of the system is:
\begin{equation}
U=E_{\mathrm{tr}}+ E_{\mathrm{rot}}+E_{\mathrm{pot}} \,,
\label{internal_energy}
\end{equation}
where
\begin{equation}
E_{\mathrm{tr}}= \sum_{i=1}^{N} \sum_{j=1}^{d} \frac{p_{ij}^{2}}{2m}
\label{kinetic_energy}
\end{equation}
is the translational kinetic energy, with $p_{ij}$ being the $j$th component of the $i$th particle's momentum $\boldsymbol{p}_{i}$, and $m$ the mass of particles.  
\begin{equation}
E_{\mathrm{rot}}= \sum_{i=1}^{N} \sum_{j=1}^{f} \frac{L_{ij}^{2}}{2I_{j}}
\label{kinetic_energy}
\end{equation}
is the rotational kinetic energy, with $L_{ij}$ being the $j$th component of the $i$th particle's generalized momentum $\boldsymbol{L}_{i}$, $I_{j}$ the corresponding principal moment of inertial, and $f=d(d-1)/2$ being the number of rotational degrees of freedom. The classical partition function of such a system is given by
\begin{equation}
\begin{aligned}
Z_{\mathrm{hd}}(N) = & \frac{1}{N!h^{dN} h^{fN}} \int  d \boldsymbol{r}^{N} \int d \boldsymbol{\omega}^{N} \int d \boldsymbol{p}^{N} \int d \boldsymbol{L}^{N} \exp(-\beta U) \\
 = & \frac{1}{N!h^{dN} h^{fN}} \int d \boldsymbol{p}^{N} \exp(-\beta E_{\mathrm{tr}})  \int d \boldsymbol{L}^{N} \exp(-\beta E_\mathrm{{rot}})  \int d \boldsymbol{r}^{N} \int d \boldsymbol{\omega}^{N} \exp[- \beta E_{\mathrm{pot}}(\boldsymbol{r}^{N}, \boldsymbol{\omega}^{N})] \\
 = & \frac{1}{N!h^{dN} h^{fN}} (\frac{2\pi m}{\beta})^{\frac{dN}{2}} [\prod_{j=1}^{f} (\frac{2 \pi I_{j}}{\beta})^{\frac{1}{2}} ]^{N} V_{\mathrm{hd}}(N).
\end{aligned}
\label{partition_function}
\end{equation}
Here $\beta=1/k_{\mathrm{B}}T$, with $k_{\mathrm{B}}$ the Boltzmann constant and $T$ the temperature. $h$ is the Planck constant. 
\begin{equation}
V_{\mathrm{hd}}(N) =  \int  d \boldsymbol{r}^{N} \int d \boldsymbol{\omega}^{N} \exp[- \beta E_{\mathrm{pot}}(\boldsymbol{r}^{N}, \boldsymbol{\omega}^{N})]
\end{equation}
is the integral over the configurational space. Because $E_{\mathrm{pot}}$ takes on values of $0$ when there is no particle overlap and $\infty$ when particle overlaps occur, $V_{\mathrm{hd}}(N)$ actually gives the free volume in the configurational space of the hard particle system. 
In the case of an ideal gas of molecules, $E_{\mathrm{pot}}$ is always zero, and particle positions and particle orientations are decoupled. This results in the free volume $V_{\mathrm{id}}= C_{\mathrm{rot}}^{N}V_{\mathrm{box}}^{N}$. Here $C_{\mathrm{rot}}$ arises from integrating the rotational degrees of freedom for a single molecule, given by $C_{\mathrm{rot}}=2 \pi^{d/2}/\Gamma(d/2)$, where it represents the solid angle subtended by the complete $(d-1)$-dimensional spherical surface of a unit sphere within $d$-dimensional Euclidean space, with $\Gamma$ being the gamma function. At a fixed temperature, both systems possess the same internal energy, given by: $U = -\frac{\partial}{\partial \beta} \ln Z(N) = \frac{1}{2}(d+f)N k_{\mathrm{B}}T$. Utilizing the relationships $F=-k_{\mathrm{B}}T \ln Z$ and $F=U-TS$, we can express the excess entropy $S_{\mathrm{ex}}(N)$ as: 
\begin{equation}
S_{\mathrm{ex}}(N)= -\frac{F_{\mathrm{ex}} }{T}= k_{\mathrm{B}}\ln{\frac{V_{\mathrm{hd}}(N)}{V_{\mathrm{id}}(N)}},
\label{S_ex}
\end{equation}
where ``ex" denotes the excess contribution beyond that of the ideal gas.

We proceed to employ Monte Carlo (MC) integration to compute the available free volume for particles confined within a periodic cubic box. Drawing inspiration from the classic example of MC integration used to estimate the area of a 2D circle, wherein a circle is inscribed within a square and the ratio of points falling inside the circle to the total number of attempted points equates to the ratio of their respective areas \cite{MC_Pi}, we adapt a similar approach. To perform this integration, we begin by placing a particle into the box with its position and orientation randomly selected. Subsequently, we proceed to randomly place a second particle. If these two particles do not overlap, we continue by introducing a third particle, and so forth. This iterative process persists until particle overlap is encountered. Let $n_{\mathrm{tot}}$ represent the total number of trials, and $n_{\mathrm{s}}(N)$ denote the count of successful trials for a given particle count $N$. The configurational free volume for a system composed of $N$ hard particles is calculated as follows \cite{Wan2018, Wan2022}:
\begin{equation}
V_{\mathrm{hd}}(N)=\frac{n_{\mathrm{s}}(N)}{ n_{\mathrm{tot}} }\times V_{\mathrm{box}}^{N}\times C_{\mathrm{rot}}^{N}.
\label{V_MC}
\end{equation}
Since the attempt to place the next particle is dependent on the successful placement of all the previous particles, Eq.~(\ref{V_MC}) can be written as
\begin{equation}
\begin{aligned}
V_{\mathrm{hd}}(N) =& \frac{n_{\mathrm{s}}(1)}{n_{\mathrm{tot}}} \cdot \frac{n_{\mathrm{s}}(2)}{n_{\mathrm{s}}(1)} \cdot \frac{n_{\mathrm{s}}(3)}{n_{\mathrm{s}}(2)} \cdots \frac{n_{\mathrm{s}}(N)}{n_{\mathrm{s}}(N-1)} \times V_{\mathrm{box}}^{N}\times C_{\mathrm{rot}}^{N} \\
          =& \prod_{i=1}^{N} P_{i} \times V_{\mathrm{box}}^{N}\times C_{\mathrm{rot}}^{N}.
\end{aligned}
\label{V_probability} 
\end{equation}
Here
\begin{equation}
P_{i}=\left\{
\begin{aligned}
&\frac{n_{\mathrm{s}}(i)}{n_{\mathrm{tot}}}=1, && i=1\\
& \frac{n_{\mathrm{s}}(i)}{n_{\mathrm{s}}(i-1)}, && 2 \le i \le N \\
\end{aligned}
\right.
\end{equation}
indicates the probability of the successful placement of the $i$th particle when there are already $(i-1)$ particles present. As the first particle can always be placed in the periodic box successfully, $P_{1}=1$. Substituting Eq.~(\ref{V_probability}) into Eq.~(\ref{S_ex}), the excess entropy $S_{\mathrm{ex}}$ is then:
\begin{equation}
S_{\mathrm{ex}} =  k_\mathrm{B} \ln (\prod_{i=1}^{N} P_{i}).
\label{S_entropy} 
\end{equation}
This is Eq.~(3) in the main text.

\end{document}